\documentclass[%
 reprint,
superscriptaddress,
 amsmath,amssymb,
 aps,
]{revtex4-1}

\usepackage{graphicx}
\usepackage{dcolumn}
\usepackage[normalem]{ulem}
\usepackage{bm}

\begin{document}

\preprint{APS/123-QED}

\title{Characterization of fracture in topology-optimized bio-inspired networks}

\author{Chantal Nguyen}
 \email{cnguyen@physics.ucsb.edu}
 \affiliation{Department of Physics, University of California, Santa Barbara, Santa Barbara, CA 93106, USA}

 \author{Darin Peetz}
\affiliation{Department of Civil and Environmental Engineering, University of Illinois at Urbana-Champaign, Urbana, IL 61801, USA}
\author{Ahmed E. Elbanna}
\affiliation{Department of Civil and Environmental Engineering, University of Illinois at Urbana-Champaign, Urbana, IL 61801, USA}
\author{Jean M. Carlson}
\affiliation{Department of Physics, University of California, Santa Barbara, Santa Barbara, CA 93106, USA}

\date{\today}

\begin{abstract}
Designing strong and robust bio-inspired structures requires an understanding of how function arises from the architecture and geometry of materials found in nature. We draw from trabecular bone, a lightweight bone tissue that exhibits a complex, anisotropic microarchitecture, to generate networked structures using multi-objective topology optimization. Starting from an identical volume, we generate multiple different models by varying the objective weights for compliance, surface area, and stability. We examine the relative effects of these objectives on how resultant models respond to simulated mechanical loading and element failure. We adapt a network-based method developed initially in the context of modeling trabecular bone to describe the topology-optimized structures with a graph theoretical framework, and we use community detection to characterize locations of fracture. This complementary combination of computational methods can provide valuable insights into the strength of bio-inspired structures and mechanisms of fracture.

\end{abstract}

\pacs{Valid PACS appear here}
\maketitle

\section{Introduction}
Understanding the relationships between architecture and function in biological materials is key to engineering bio-inspired structures for strength and resilience. Materials found in nature must be spatially arranged to withstand repeated loading while facilitating various biological functions. In this paper, we use multi-objective topology optimization, finite element modeling, and network science methods to generate and analyze a range of structures with varying emphases placed on maximizing stiffness, perimeter, and stability. We explore how differently weighting these objectives influences robustness and resistance of these structures to failure.

The bio-inspired structures we develop in this paper are motivated by the challenge of reverse-engineering trabecular bone, a type of bone tissue that consists of an interconnected network of small struts called trabeculae. Its porous structure allows it to be lightweight, though it is weaker than the other type of bone tissue, cortical bone, which is hard, dense, and shell-like. Trabecular bone has roughly ten times the surface area of cortical bone. The pores in trabecular bone hold bone marrow, nerves, and blood vessels, and the increased surface area facilitates bone resorption and remodeling. This tradeoff between the pore distribution and strength drives our choice of objectives in constructing structures guided by the emergent properties of vertebral trabecular bone.

Continuum topology optimization is a method that, given a set of objectives and constraints, optimizes the distribution of material within a domain \cite{topopt_bendsoe_sigmund}. We are motivated to use topology optimization to generate bone-inspired structures by the premise of Wolff's law \cite{wolff}. Wolff's law states that, over time, trabecular bone remodels its architecture to adapt to the loads it is regularly subjected to. That is, it will `self-optimize' itself into a structure that is more stiff along the primary loading directions. Analogously, multi-objective topology optimization starts from an initial density distribution, applies specified loads that in our case represent uniaxial loading in vertebrae, and minimizes a weighted sum of objective functions to achieve a desired architecture. Here, the objective functions represent compliance (inverse stiffness), perimeter (the 2-D analog of surface area), and stability. Conceptually speaking, we assume that real bone is the outcome of a biological optimization procedure, but the quantities being optimized are unknown. While the topology-optimized structures are not intended to mimic bone, in isolating material properties associated with bone and varying the weights of corresponding objective functions, we examine how the relative weighting impacts overall toughness and robustness to failure.

The topology-optimized structures are disordered planar networks. We extract from them graph models consisting of edges representing struts (trabeculae), joined together at nodes that correspond to the branch points where the struts meet. This allows us to extract topological metrics that quantify the architecture of the network. This network-based method adapts the modeling approach developed by Mondal et al. \cite{avik} which modeled real human trabecular bone from micro-CT images.

We analyze the mechanical response of the topology-optimized networks by converting the networks to finite element models in which each edge is represented by a beam. We simulate compressive loading and failure in the beam-element models, and we investigate mechanics at scales ranging from individual beams to the entire network. In combining these computational methods, many of which have seen limited application to trabecular bone and bone-inspired materials, we relate the mechanics of bone-like structures to their architecture and identify how topology informs fracture. Our results inform the development and design of bio-inspired networked structures that are robust and strong.

\section{Multi-objective topology optimization}

The topology optimization process begins by assuming an initial two-dimensional density distribution on a discretized uniform grid of elements, then iteratively 1) performs a finite element analysis step that simulates mechanical deformation, 2) carries out a gradient-based optimization step that updates the density distribution, and 3) evaluates the objective until convergence \cite{peetz_thesis}. Three objectives were used: compliance (inverse stiffness) minimization, perimeter maximization, and stability maximization. The objective functions are combined as a weighted sum to form a single objective function that is evaluated in the iterative optimization procedure. Adjusting the weights of each objective function can result in highly variable topologies.

Each element has a density that can take on any value between 0 (void) and 1 (solid), but intermediate values are penalized using the solid isotropic material with penalization model (SIMP) \cite{topopt_bendsoe_sigmund} to ensure that the result contains binary density values. We include an area constraint in the optimization problem so that the total area of each generated structure is effectively constant. While the topology optimization method developed here is limited to 2-dimensional structures, it can be generalized to three dimensions, albeit with a higher computational cost. 

\begin{figure}[t!]
\includegraphics[width=\linewidth]{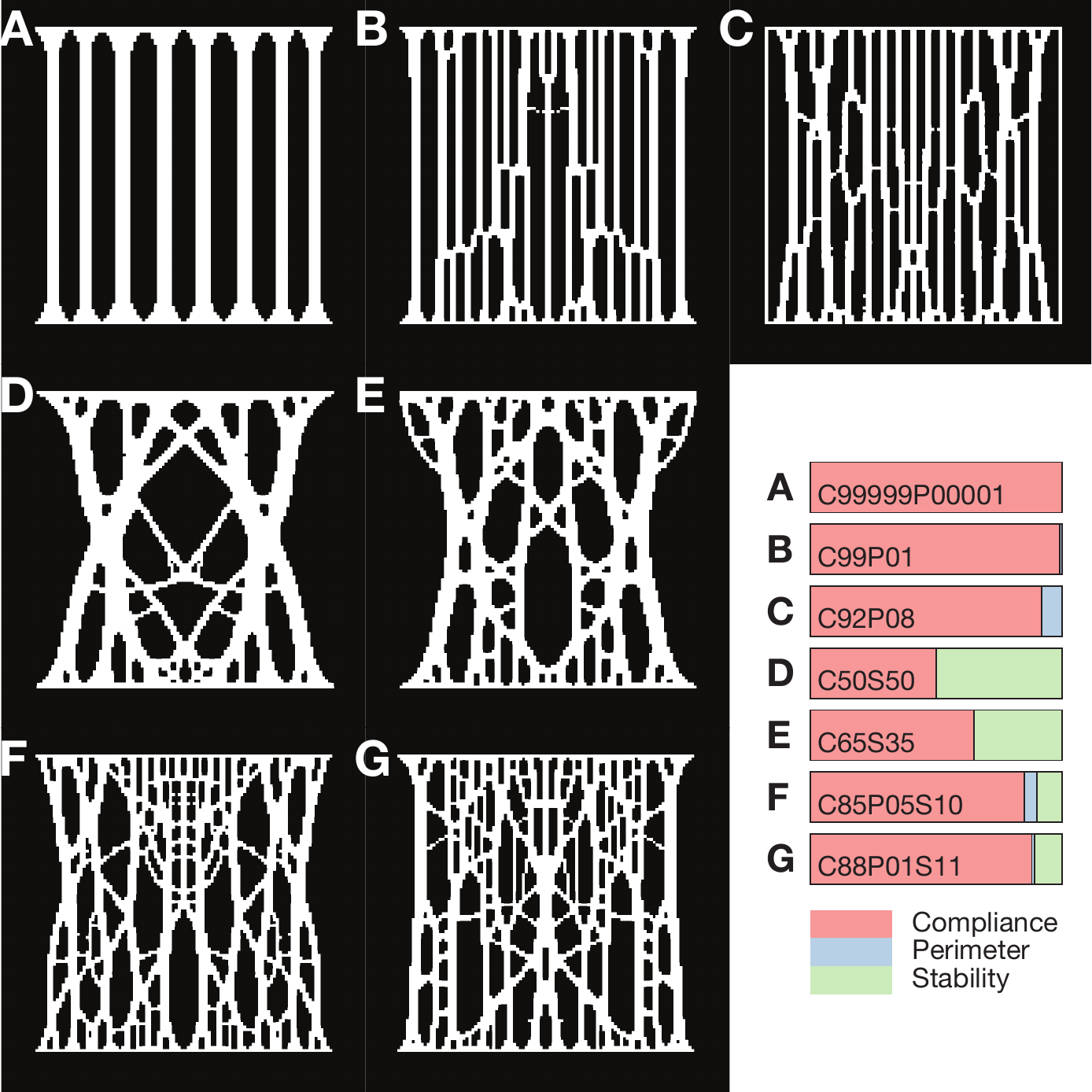}
\caption{Example 2-D topology-optimized structures generated by varying objective weights. The horizontal bar plot in the lower right shows the relative weights assigned to the compliance, perimeter, and stability objectives for each image. Weights sum to one. Panels A-G: C99999P00001, C99P01, C92P08, C50S50, C65S35, C85P05S10, and C88P01S11, respectively. A total of 12 structures were generated for each of the seven parameter sets shown here; all structures for each parameter set are shown in the Supplemental Material.}
\label{fig:topopt}
\end{figure}

The most basic topology optimization problem is that of minimizing compliance (weights of perimeter and stability functions are set to zero) with an area constraint. The topology optimization problem for minimization of compliance $C$, with a constraint on the area fraction, is conventionally defined as
\begin{align}
   \min_{\rho}\  & C = \mathbf{u}^T\mathbf{K}\mathbf{u}, \label{eq:compliance}\\
   \text{s.t.}\  & \frac{1}{A_\Omega} \sum_{e=1}^N \rho_e A_e \leq A, \nonumber
\end{align}
where $\mathbf{K}$ is the material stiffness matrix, $\mathbf{u}$ is the vector of displacements, $A_\Omega$ is the total area of the domain, $\rho_e$ is the density of element $e$, $A_e$ is the area of each element, and $A$ is a specified total area fraction. Here, $\mathbf{u}$ is related to the vector of applied loads, $\mathbf{f}$, through the relation
\begin{equation}
    \mathbf{K}\mathbf{u} = \mathbf{f}.
\end{equation}
Compliance is minimized, or equivalently, stiffness maximized, to minimize the displacement undergone by the structure in response to loading. Minimizing compliance alone produces a structure primarily consisting of thick rods aligned with the principal direction of loading (Fig. \ref{fig:topopt}A). Hence, an anisotropic architecture can give rise to increased stiffness when the elements (trabeculae) are preferentially aligned with the loading direction. 

However, trabecular bone does not consist of thick parallel rods. The surface of trabecular bone is necessary for its remodeling cycle, which requires contact with surrounding bone marrow for new osteoclasts to form \cite{seeman_bone_review}. Bone is resorbed by osteoclasts, with new bone deposited on the surface by osteoblasts. Trabecular bone also has a much higher surface area compared to cortical bone and consequently a large number of pores that hold marrow, nerves, and blood vessels.

Reverse-engineering trabecular bone to produce a structure of similar flexibility and lightness will require taking perimeter into account as in the objective function. Here we define $P$, the perimeter (2-D) or surface area (3-D) of the structure, in a dimension-agnostic form as
\begin{equation}
    \max_{\rho}\  P = \int \Delta \rho\ \text{d}\Omega,
       \label{eq:perimeter}
\end{equation}
where $\rho$ is the material density or volume at any point in the structure. Numerically, this translates to a sum of density changes across all element boundaries.
Setting the perimeter function weight to a non-zero value and optimizing for both compliance and perimeter, while keeping the same volume, results in a structure with a greater number of thinner struts, rather than fewer, thicker ones. Most of these thin struts are aligned in the principal loading direction, while a few are transverse.

Previous studies applying topology optimization to explore trabecular bone structure have considered only compliance as an objective function and included a perimeter constraint \cite{boyle_kim_topopt,jang_kim_topopt}. However, depending on the weights used, including only compliance (and perimeter) objective functions can result in an unstable model, such as one that consists of long, thin vertical rods. The instability of this model is represented by its critical buckling load, $P_{crit} = \max_{i = 1, ..., N_{dof}} P_i$. The objective in this case is to maximize the critical buckling load defined by the generalized eigenvalue equation
\begin{equation}
\left[\mathbf{G(u)} - \frac{1}{P_i} \mathbf{K}\right]\mathbf{\Phi_i} = 0, \hspace{5mm} i = 1, \hdots , N_{dof},
\end{equation}
where $\mathbf{G(u)}$ is the geometric stiffness matrix and $\mathbf{\Phi_i}$ is the eigenvector associated with the $i$th buckling load. To avoid degeneracy of the eigenvalues $1/P_{i}$, which can result in poor or incorrect convergence of the optimizer, we apply a bound formulation \cite{topopt_bendsoe_sigmund} such that the stability optimization problem is written as 
\begin{align}
   \min_{\rho}\  & \beta, \\
   \text{s.t.}\  & \alpha^i \left(\frac{1}{P_i}\right) \leq \beta, \ i = 1, \hdots , N_{dof}, \ \nonumber \\
   & \left[\mathbf{G(u)} - \frac{1}{P_i} \mathbf{K}\right]\mathbf{\Phi_i} = 0, \ i = 1, \hdots , N_{dof}, \nonumber
      \label{eq:stability}
\end{align}
where $\alpha$ is a number slightly less than 1, e.g. 0.95, which ensures that each eigenvalue is slightly larger than the next. Note that this bound formulation will only actively impact eigenvalues near one end of the spectrum and eigenvalues in the interior or near the other end of the spectrum will inherently satisfy the constraint. As a result, we can safely truncate the series from $N_{dof}$ terms to a much smaller number such as $n=10$. Optimizing for stability as well as compliance and perimeter further increases the number of struts as well as those oriented at a nonzero angle to the primary loading (vertical) direction.

The multiple objectives are combined as a weighted sum, where the weights can be varied to change the relative importance of each objective:
\begin{align}
   \min_{\rho}\ & w_1 C_0 - w_2 P_0 + w_3 \beta_0, \\
   \text{s.t.}\  & \alpha^i \left(\frac{1}{P_i}\right) \leq \beta, \ i = 1, \hdots , N_{dof}, \ \nonumber \\
   & \left[\mathbf{G(u)} - \frac{1}{P_i} \mathbf{K}\right]\mathbf{\Phi_i} = 0, \ i = 1, \hdots , N_{dof}, \nonumber \\
   & \frac{1}{A_\Omega} \sum_{e=1}^N \rho_e A_e, \nonumber \\
   & \sum_{i=1}^3 w_i = 1, \nonumber
\end{align}
where $w_i$ are the respective weights on each of the objective functions $C_0$, $P_0$, and $\beta_0$, which refer to normalized compliance, perimeter, and stability, respectively (Eqs. \ref{eq:compliance}, \ref{eq:perimeter}, and \ref{eq:stability}). Here we normalize by independently optimizing for each of the objectives separately and then evaluating each objective function on each optimized structure. The functions are then normalized relative to the maximum and minimum values across each of the structures.

Note that the purpose of normalization is to make the magnitude of each function more consistent. As a result, the actual values of the function weights for one system are somewhat arbitrary in that they depend on the normalization procedure used. As such, the weights are only truly meaningful when compared relative to each other and/or across different optimization problems.

We generate topology-optimized structures for a total of seven different sets of objective weights. One example structure for each parameter set is shown in Fig. \ref{fig:topopt}; all remaining structures are included in the Supplemental Material. Each set contains twelve different structures. Each structure is generated from the same initial density distribution, with a small perturbation added to ensure that each optimization with the same weights will converge to a different structure. We label each set of structures with the letters C, P, and/or S, representing compliance, perimeter, and stability objectives, respectively, followed by the corresponding weight (times 100) of the objective function used to generate the structures. 

Fig. \ref{fig:topopt}A is an example structure from the set labeled C99999P00001, which is representative of optimizing all but entirely for compliance. The weight of the compliance function is 0.99999, rather than 1 even. If the compliance weight were 1, for some initial conditions, it is possible that the result would be a contiguous piece of material with no porosity. Hence, we assign a very small weight of 0.00001 to the perimeter objective; combined with the different initial conditions, this promotes variation in topology. Stability is not considered in this case.

Figs. \ref{fig:topopt}B-C, labeled C99P01 and C92P08, respectively, are generated by including weights for both compliance and perimeter, resulting in an increased number of thinner struts and consequently a greater number of pores.

Figs. \ref{fig:topopt}D-E, labeled C50S50 and C65S35, respectively, are generated by including weights for compliance and stability, but omitting the perimeter objective. The resulting structures consist of much thicker struts that are largely oriented at an angle to the vertical. The structures are also noticeably concave at each side.

Figs. \ref{fig:topopt}F-G, labeled C85P05S10 and C88P01S11, respectively, are generated from combining all three objectives. These structures contain more struts and small pores than the other sets, with a few longer vertical columns joined by a number of shorter angled elements.

\begin{figure*}[]
\includegraphics[width=\textwidth]{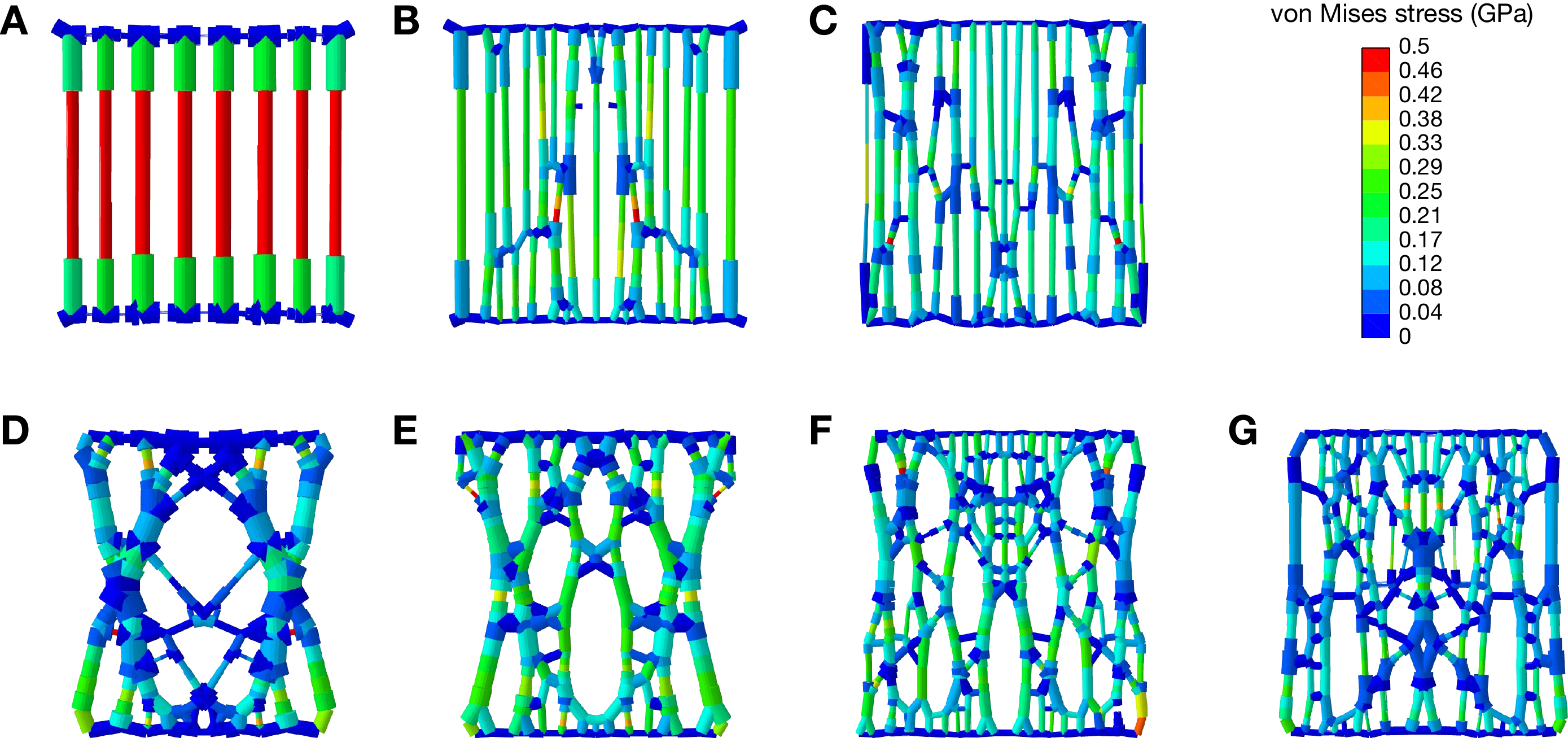}
\caption{Example beam element models. Color of beams represents spatial distribution of von Mises stress in example structures for each parameter set. Each model is shown at the timestep immediately preceding the first element failure in each respective simulation. A: C99999P00001; B: C99P01; C: C92P08; D: C50S50; E: C65S35; F: C85P05S10; G: C88P01S11.}
\label{fig:topopt_stress}
\end{figure*}

\begin{figure*}[]
    \includegraphics[width=\linewidth]{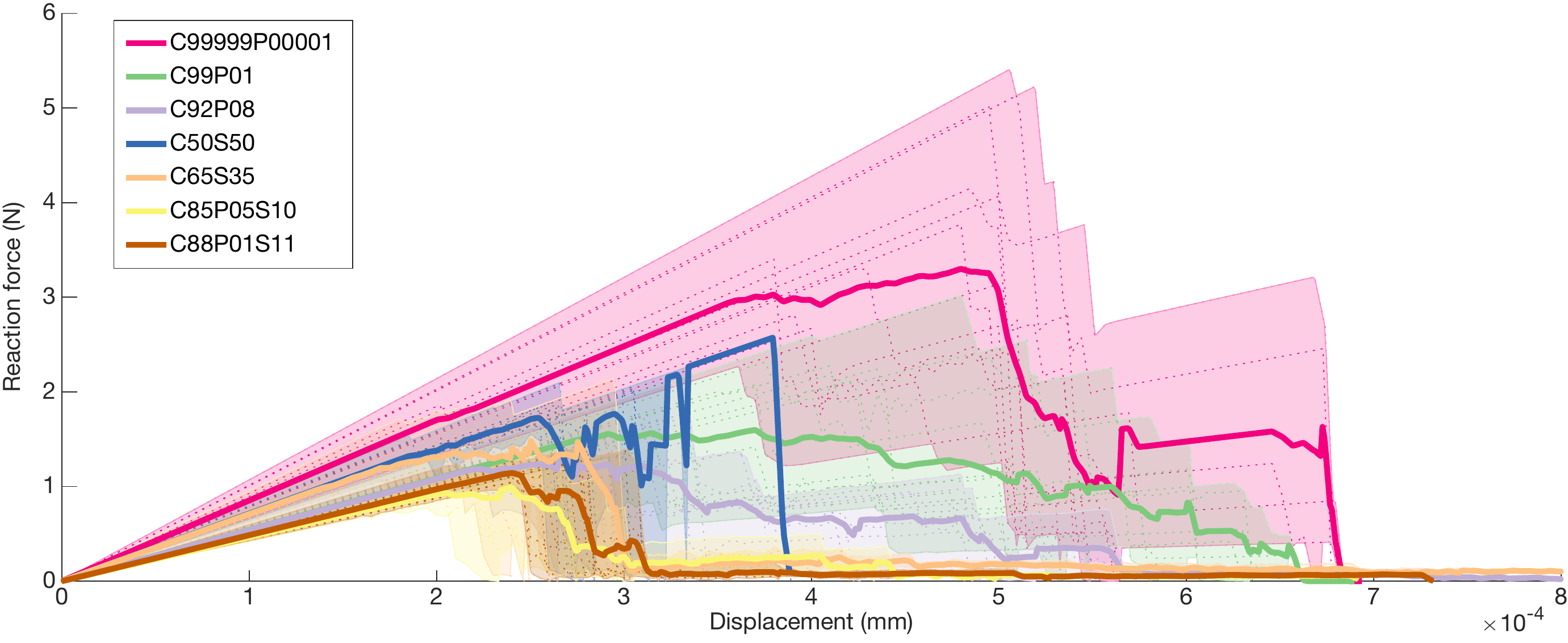}
    \caption{Force-displacement response. The force-displacement curve for each structure is indicated by a thin dashed line; the average curve for each parameter set is shown as a thick solid line. Shaded areas represent the regions spanned by the highest and lowest reaction force for each parameter set. }
    \label{fig:forcedisp}
\end{figure*}

\section{Network modeling and mechanical simulation}

\subsection{Skeletonization}
From topology-optimized images, we generate graph models, following \cite{avik}, that allow us to utilize existing graph theoretical methods to efficiently analyze the topology of networked structures. Converting a topology-optimized structure to a graph begins with skeletonization: the ``skeleton'' of each image is determined by progressively thinning the image until its medial axis, a one-pixel-wide line running through the center of the network, is found. This medial axis, or skeleton, is then converted to a graph by setting nodes at branch points where 3 or more struts meet, with edges corresponding to struts themselves. The edges are weighted according to the respective average thicknesses of corresponding struts. Skeletonization and graph conversion are accomplished using the Skeleton3D and Skel2Graph toolboxes for MATLAB \cite{osteocyte_kerschnitzki}. Strut thicknesses are computed using the BoneJ plug-in \cite{bonej} for ImageJ (National Institutes of Health, Bethesda, MD).

\subsection{Beam element models}
To simulate mechanical loading and deformation, we translate these graphs into streamlined finite element models. Rather than meshing the trabecular model, we generate beam-element models from the graphs, where each link is represented by a Timoshenko beam with a uniform thickness corresponding to its weight (Fig. \ref{fig:topopt_stress}). Nodes in the beam-element model correspond directly to nodes in the network.

Mechanical loading is simulated with Abaqus FEA (Dassault Syst\`{e}mes, V\'{e}lizy-Villacoublay, France). The beam-element model is compressed from the top and bottom, representing loading along the superior-inferior direction, the primary loading axis in vertebrae. The von Mises stress at each link is computed at each time step, along with the force and displacement of each node. 

We solve the models in the linear-elastic regime, where the stress is linear as a function of strain. We also model failure by setting von Mises stress as a failure criterion; when the stress in a beam reaches the critical stress value, the beam is said to have failed and is removed from the simulation. We arbitrarily set the failure criterion to be a von Mises stress of 0.5 MPa; as the response is linear, this value can be scaled up or down with no qualitative change in the overall behavior.

We note that the skeletonization and network conversion process is limited by its inability to fully capture non-uniform trabecular thicknesses or increased bulk at branch points (nodes). This tradeoff, however, greatly simplifies modeling and provides a streamlined approach to relating topology with mechanics. To improve the resolution of trabecular thickness in beams with nonuniform widths, we divide longer beams into five segments, such that each segment can have a different thickness.

\subsection{Bulk force-displacement response}
Force-displacement curves for the seven beam-element models generated from the topology-optimized structures (Fig. \ref{fig:topopt}) are compared in Fig. \ref{fig:forcedisp}. 
We model the structures in the linear-elastic regime with a von Mises stress failure criterion. The force-displacement curves are hence linear until the initialization of beam failure, whereupon they exhibit large decreases until reaching zero, at which point the structure is said to have failed completely. The force-displacement response after this point exhibits fluctuations that are artifacts of wave propagation in the simulation and are not considered in the analysis of the results. The curves in Fig. \ref{fig:forcedisp} are truncated where the reaction force reaches zero, and the full force-displacement curves for each model are included in the Supplemental Material. 

\begin{figure*}[]
    \includegraphics[width=\linewidth]{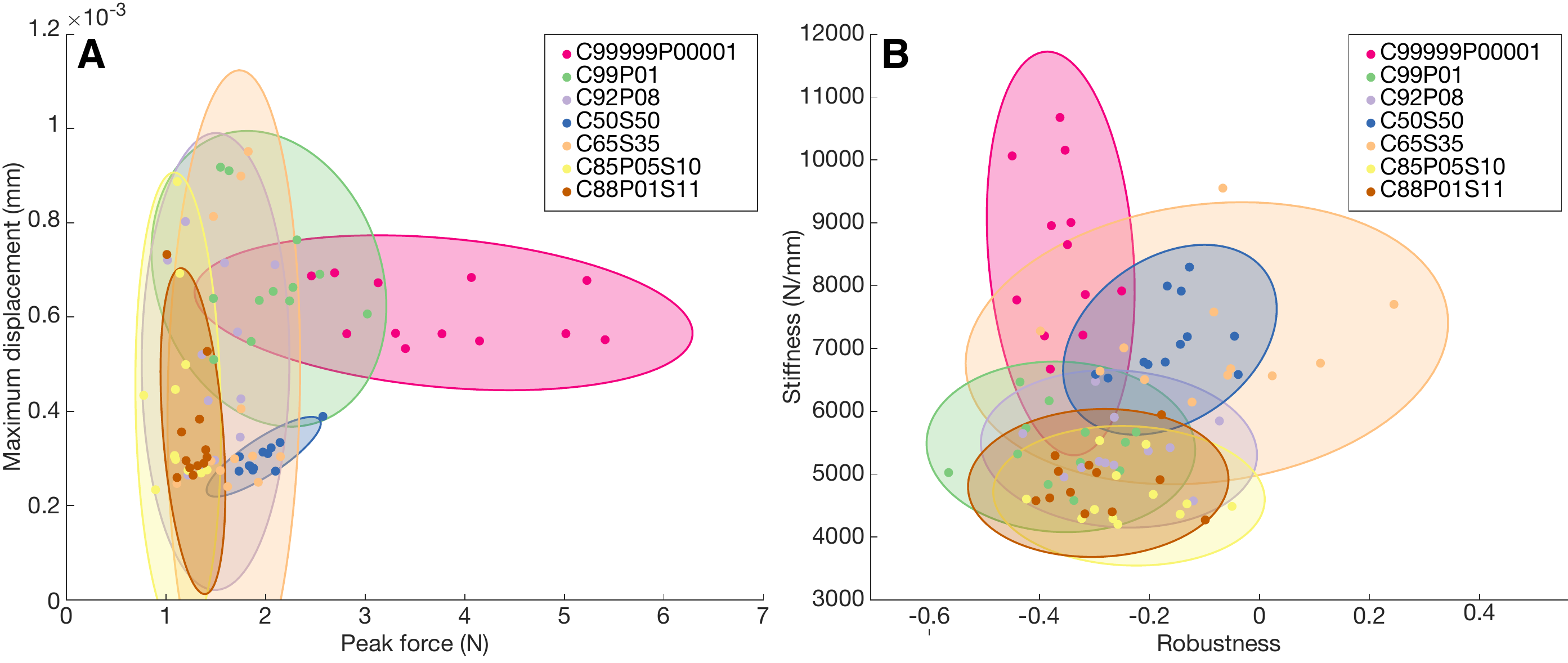}
    \caption{Ashby plots comparing properties of different optimization parameter sets. Panel A compares the maximum displacement before complete failure with the peak reaction force attained. Panel B compares stiffness, the slope of the force-displacement curve in the linear regime prior to failure, with robustness, measured as the relative change between the peak forces of the original and perturbed models. Shaded ellipses represent 2$\sigma$ confidence intervals.}
    \label{fig:ashby}
\end{figure*}

On average, stiffness (the slope of the force-displacement curve in the initial linear regime) is greatest for C99999P00001, the parameter set for which compliance minimization was most highly weighted. However, C50S50 and C65S35 demonstrate slightly higher average stiffness than C99P01 and C92P08, which have greater compliance minimization weights. The models with lowest stiffness are C85P05S10 and C88P01S11.

We use two additional metrics to quantify mechanical response: the peak reaction force typically attained at the onset of element failure, and the maximum displacement at total system failure (when the reaction force reaches 0). The peak force represents the strength of the model, while the maximum displacement serves as a proxy for the ductility of the structure as it undergoes fracture. A large maximum displacement could indicate that stresses redistribute such that the entire structure does not fail immediately when the first failure occurs. The distributions of peak force and maximum displacement are compared in an Ashby plot in Fig. \ref{fig:ashby}A. The highest peak forces are given by C99999P00001, followed by C99P01, while the peak force for the other parameter sets are comparable. The maximum displacement varies greatly for some parameter sets, in particular C92P08, C65S35, C85P05S10, and C88P01S11, while the variation in displacement is considerably smaller for C99999P00001 and C50S50.

We note that while C99999P00001 demonstrates the highest peak forces, it also has the largest variation in peak force. Hence, slight variations in structure across models, despite being generated under the same optimization criteria, can result in significantly different mechanical response. To probe robustness, we perturb each structure slightly and subject them to the same loading conditions as the original models. For each model, each node is shifted in both $x$- and $y$- coordinates by a small random distance of order 1\% of the length of the structure. 

For the purposes of this paper, we define robustness as the relative change in peak force between the original and perturbed models: $(F_{\text{peak, original}} - F_{\text{peak, perturbed}})/ F_{\text{peak, original}}$. Robustness is plotted against the stiffness of the original model in Fig. \ref{fig:ashby}. In some cases, the perturbed model can exhibit a greater peak force than the original model, indicated by a positive robustness score. We observe that C99999P00001, which demonstrated the greatest variation in peak force among original models, exhibits relatively low robustness, with large spread in stiffness values. C65S35 exhibits the greatest variation in robustness, with several instances in which the perturbed model was stronger than the original model. C50S50 shows slightly lower robustness than C65S35; C50S50 and C65S35 exhibit roughly similar stiffness values and are the second stiffest models after C99999P00001. C99P01, C92P08, C85P05S10, and C88P01S11 demonstrate similar stiffness and robustness. 

These results suggest that while assigning almost all weight to compliance minimization can produce structures that are on average stiffer and tougher, these structures can be prone to small perturbations in geometry. Moreover, optimizing for compliance and perimeter without accounting for stability can result in structures that are less robust and less stiff than those generated by assigning considerable weight to stability maximization. However, structures with small weights on both perimeter and stability objectives remain weaker and less robust than those for which perimeter is not considered.

\subsection{Stress distribution}

\begin{figure*}[]
\includegraphics[width=\textwidth]{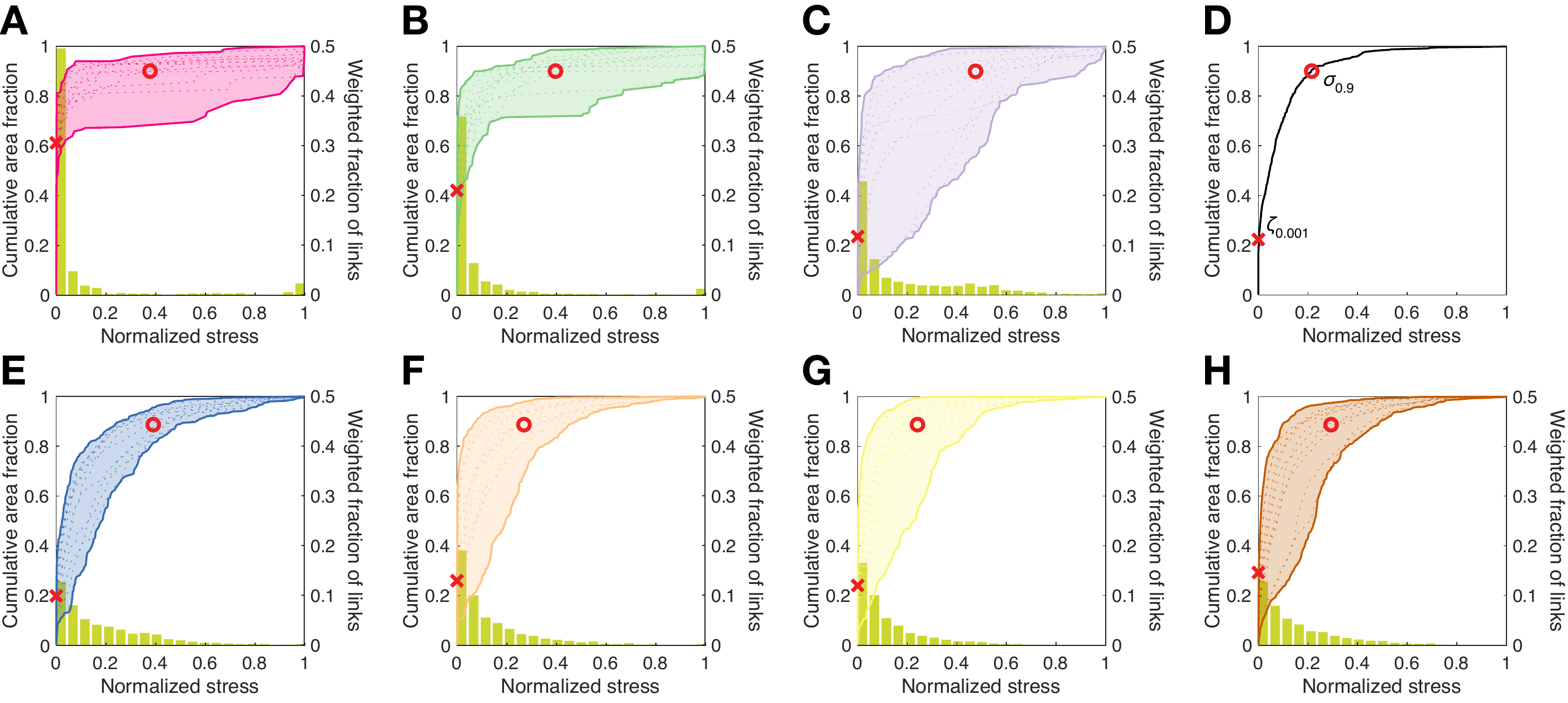}
\caption{Stress distributions. A: C99999P00001; B: C99P01; C: C92P08; D: example cumulative stress distribution; E: C50S50; F: C65S35; G: C85P05S10; H: C88P01S11. Histograms represent the average distribution of normalized stress for each parameter set, weighted by the thickness of each link. The shaded regions illustrate the variation in the cumulative distribution of normalized stress, expressed in terms of the fraction of area occupied by the links (normalized by the area of the entire model). Dotted lines within the shaded regions correspond to the distributions of each individual model. Red crosses represent average $\zeta_{0.001}$ and $\sigma_{0.9}$ for each parameter set, as illustrated by the example in the panel D.}
\label{fig:stressfrac}
\end{figure*}

The fragility of these structures may be linked to the spatial distribution of stress: whether the stress is distributed relatively evenly or concentrated in a few beams. The distribution of (von Mises) stress across beams can vary greatly between parameter sets, as visualized in Fig. \ref{fig:topopt_stress}. Fig. \ref{fig:stressfrac} illustrates the distribution of stress, normalized to the highest stress value in one beam in each model, averaged over all models in a set (histogram). In the models without stability objectives (top row), a large area fraction exhibits no stress, demonstrated by a considerable peak at 0. The distribution for C99999P00001, however, shows that in some models, a small fraction of links bears almost all of the stress. In contrast, the models with stability objectives (bottom row) demonstrate a peak at 0 with relatively heavy tails. 

\begin{table}
\begin{tabular}{r|c|c}
Set & $\zeta_{0.001}$ & $\sigma_{0.9}$ \\ \hline
C99999P00001 & 0.612 & 0.378 \\ 
C99P01 & 0.421 & 0.397 \\ 
C92P08 & 0.236 & 0.475 \\ 
C50S50 & 0.199 & 0.391 \\ 
C65S35 & 0.260 & 0.270 \\ 
C85P05S10 & 0.187 & 0.241 \\ 
C88P01S11 & 0.162 & 0.293 \\
\end{tabular}
\caption{Average $\zeta_{0.001}$ and $\sigma_{0.9}$ values for each set. $\zeta_{0.001}$ gives the fraction of beams with normalized stress less than or equal to 0.001, and $\sigma_{0.9}$ gives the normalized stress value wherein 90\% of beams bear stress less than equal to this value. Stress is normalized to the largest stress value in a single beam in each individual structure.}
\label{tab:zeta}
\end{table} 

Fig. \ref{fig:stressfrac} also shows the cumulative fraction of beams that bear normalized stress values between 0 and 1 (colored shaded regions). For C99999P00001, and to a lesser extent, C99P01, a notable fraction of beams have normalized stress close to 0. Their cumulative distributions rise sharply compared to those with stability objectives before flattening out. To quantify the stress distribution, we compute two metrics $\zeta_{0.001}$ and $\sigma_{0.9}$. $\zeta_{0.001}$ is the fraction of total area with normalized stress less than or equal to 0.001, and $\sigma_{0.9}$ is the normalized stress value such that 90\% of the total area bears stress less than or equal to this value; similar metrics were previously defined in the context of trabecular bone in \cite{avik}. Average values for $\zeta_{0.001}$ and $\sigma_{0.9}$ are tabulated in Table \ref{tab:zeta}. $\zeta_{0.001}$ is highest for C99999P00001; approximately 61\% of the total area -- corresponding to 67\% of beams -- bear almost no stress, followed by C99P01 at 42\% (52\% of beams). For the remaining models, which all include stability weights except for C92P08, $\zeta_{0.001}$ is lower, representing between 16\% and 26\% of area that is unstressed, indicating that stress is distributed more evenly for these models. 

For $\sigma_{0.9}$, the highest values are found for the three models with the highest compliance weights. These models have relatively high $\zeta_{0.001}$ values as well, thus containing a larger percentage of low-stress area with the stress more evenly distributed on the remaining elements. $\sigma_{0.9}$ is moreover relatively high for C50S50, which also has a low $\zeta_{0.001}$ value, indicating that the stress distribution is less skewed. Overall, $\sigma_{0.9}$ ranges between 0.24 and 0.47 for all models, implying that a small percentage of beams bear large stresses.

The models with stability objectives are most similar in visual resemblance to trabecular bone, and the shape of their stress distributions is also the most similar to that of bone \cite{avik}. For the models with stability objectives, however, $\zeta_{0.001}$ remains much lower than for bone, which is on average approximately 0.43 \cite{avik}, while this value is surpassed for C99999P00001 and C99P01. For bone, approximately 6.7\% of the total volume fraction bears less than 90\% of the normalized stress \footnote{Note that the average values of $\zeta_{0.001}$ and $\sigma_{0.9}$ as reported in \cite{avik} are 0.410 and 0.136, which were determined with respect to a cumulative distribution over the number of links, rather than volume fraction. As each link has a different volume, these values correspond to 0.428 and 0.067, respectively, when volume fraction is taken into account.}, indicating that the stress distributions are considerably less skewed for the topology-optimized models than for bone -- note, however, that the topology-optimized structures generated here are two-dimensional, while the bone volumes analyzed previously are three-dimensional.

\subsection{Community detection}
We use community detection to investigate whether the topology of the network encodes information about likely points of failure. We observe that locations of failure -- i.e., the most stressed beams in the finite element models -- do not generally correspond with the thinnest elements, and there is no preferred orientation associated with the failed beams. We hypothesize that elements corresponding to links that connect two different communities -- ``boundary links'' -- are more likely to fail than elements within a community.

Community detection is a method of determining clusters (communities) that contain dense within-cluster connections, with sparse connections to the rest of the network \cite{newman_book}. The development of community detection algorithms and their application as a beginning phase of network structure or function diagnostics is a focus of network science \cite{communitydetection_review}.
Community detection has been used to characterize social interactions, brain function, and much more, but most pertinently to characterize force chains in granular materials \cite{community_detection_granular,community_detection_granular2}. Granular packings have been described by assigning nodes to individual particles and edges to contact forces between particles \cite{particles_grains}. Community detection can extract information about force chains, networks that typically resemble interconnected filaments primarily aligned with the principal axes of loading. 

Here, we perform community detection to identify whether failure locations reside in any particular locations within the network topology. Community detection typically involves maximizing a modularity function $Q$ that identifies community structure relative to a null model $P$ \cite{newman_book,particles_grains}:
\begin{equation}
Q  = \sum_{ij} [W_{ij} - \gamma P_{ij}]\delta(g_i,g_j),
\end{equation}
where $W_{ij}$ is the weight of the edge between nodes $i$ and $j$, $\gamma$ is a resolution parameter that controls community size, $P_{ij}$ specifies the expected weight of the edge between nodes $i$ and $j$ under the null model, $g_i$ is the community assignment of node $i$, and $\delta(g_i, g_j)$ is the Kronecker delta.

The null model is commonly chosen to be a random rewiring of nodes with the degree distribution kept constant (Newman-Girvan null model):
\begin{equation}
P_{ij} = \frac{s_i s_j}{2m},
\end{equation}
where $s_i$ is the weighted degree of node $i$ and $m$ is the sum of all edge weights in the network (i.e., $m = \frac{1}{2} \sum_{ij} W_{ij}$).
This null model assumes that connections between any pair of nodes is possible. However, because the networks are spatially embedded, and long-range connections that span large spatial distances are impossible, we choose a \textit{geographical null model}, initially developed for use in the study of brain networks and subsequently adapted for granular networks \cite{community_detection_granular}:
\begin{equation}
    P_{ij} = \rho B_{ij},
\end{equation}
where $\rho$ is the mean edge weight of the network and $\mathbf{B}$ is the binary adjacency matrix of the network (i.e., the adjacency matrix where all nonzero edge weights have been set to 1).

\begin{figure*}[]
    \includegraphics[width=\linewidth]{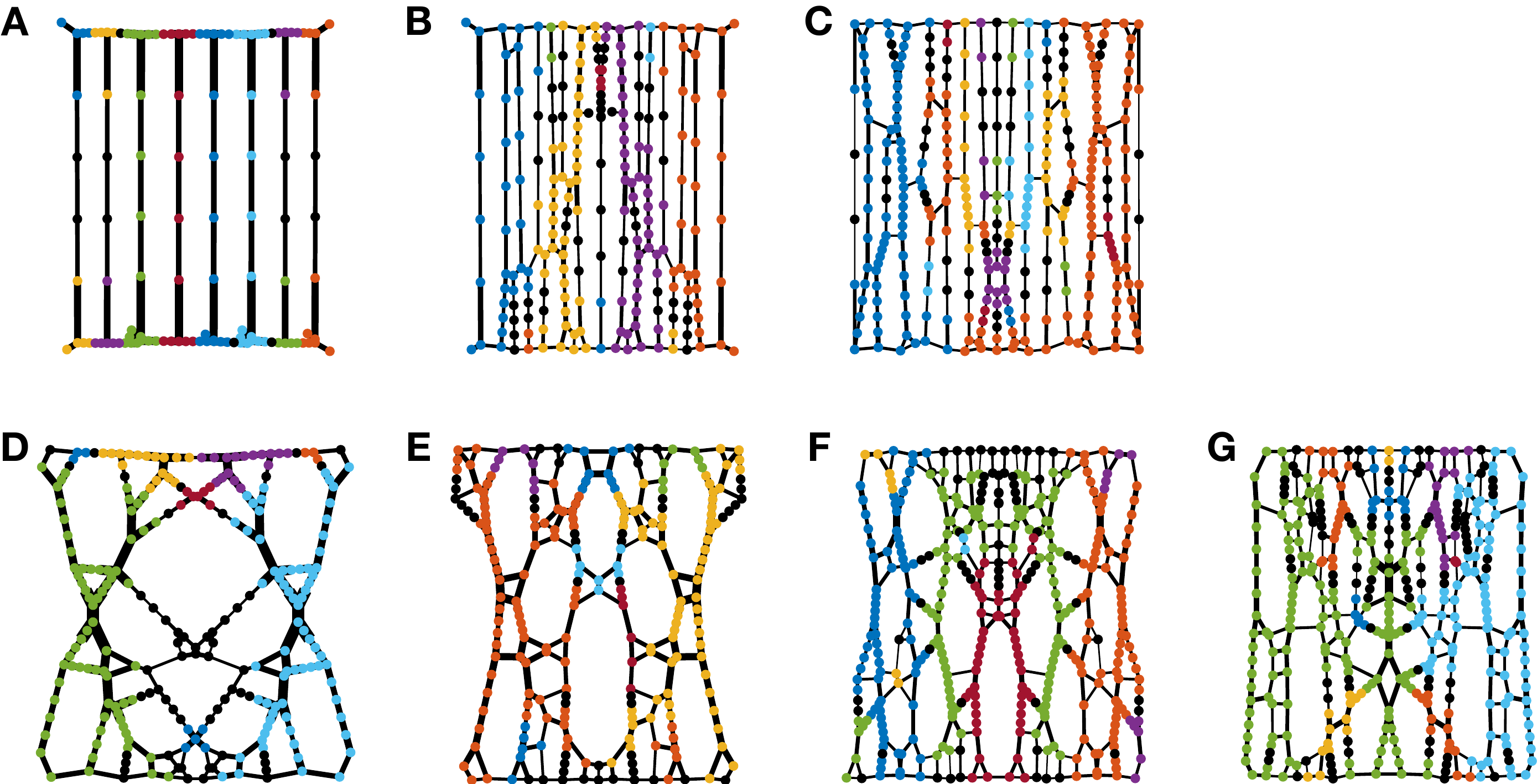}
    \caption{Example of community structure for each parameter set. A: C99999P00001, B: C99P01, C: C92P08, D: C50S50, E: C65S35, F: C85P05S10, G: C88P01S11. Nodes are colored to distinguish between communities. Black nodes represent communities of one node.}
    \label{fig:nullmodels}
\end{figure*}

The geographical null model produces communities that are anisotropically aligned with the vertical direction and thus reminiscent of force chains. The resolution parameter $\gamma$ modulates the size and number of communities. We set $\gamma$ to 1.6. Examples of community structure are shown in Fig. \ref{fig:nullmodels}.

We observe that failures tend to occur at the boundaries between communities, i.e., in links that connect two different communities. We quantify statistical significance with the Bayes factor, which represents the inverse of the ratio of probability of the data given the null hypothesis -- that the probability $q$ of a failure occurring at a boundary link is equal to the fraction of boundary links in the network $l_{bd}/L$ -- to the probability of the data given the alternative hypothesis -- that the probability $q$ of failure occurring at a boundary link is unknown and where we assume a uniform prior on $[0,1]$. The Bayes factor is given by
\begin{equation}
    BF = \frac{P(F_{bd} = f | F_{tot}, q\ \text{unknown})}{P(F_{bd} = f | F_{tot}, q = l_{bd}/L)}, 
\end{equation}
where $F_{bd}$ is the number of failures at boundaries, $F_{tot}$ is the total number of failures, $l_{bd}$ is the total number of boundary links, and $L$ is the total number of links. Furthermore, 
\begin{align}
    P(&F_{bd} = f | F_{tot}, q = l_{bd}/L) \\
    &= {F_{tot} \choose f} (l_{bd}/L)^{f}(1-l_{bd}/L)^{F_{tot}-f},
\end{align}
and
\begin{align}
    P(&F_{bd} = f | F_{tot}, q\ \text{unknown}) \\
    &= {F_{tot} \choose f} \int_0^1 q^f (1-q)^{F_{tot}-f}\\
    &= {F_{tot} \choose f} \text{B}(f+1,F_{tot}-f+1),
\end{align}
where B is the beta function.
Then the Bayes factor is given by
\begin{align}
    BF = \frac{\text{B}(f+1,F_{tot}-f+1)}{(l_{bd}/L)^{f}(1-l_{bd}/L)^{F_{tot}-f}}.
\end{align}
If $BF > 10^2$, or similarly $\ln BF > 5$, then the evidence strongly supports the alternative hypothesis over the null hypothesis.

We find that the fraction of failures that occur at these boundary links ranges between 0.58 and 0.73 for structures in sets C50S50, C65S35, C85P05S10, and C88P01S11. The fractions are smaller for the sets without stability objectives, and decreases as the compliance weight increases. In contrast, the fraction of links in the networks that are boundary links ranges between 0.25 and 0.32. 

\begin{figure*}[]
    \includegraphics[width=\textwidth]{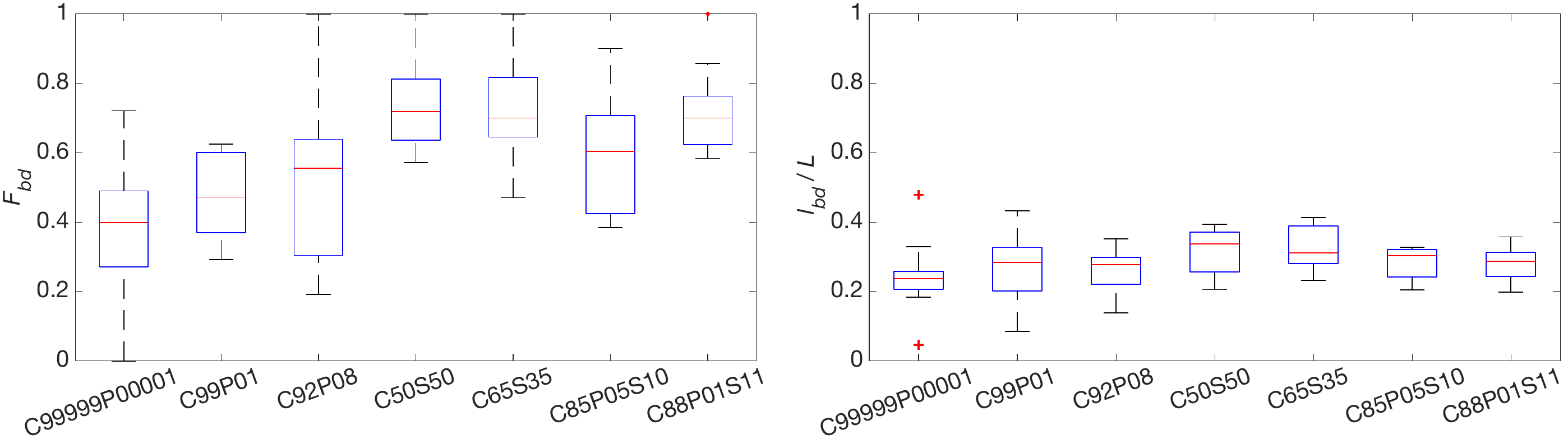}
    \caption{Variation in fraction of failures that occur at boundaries between communities ($F_{bd}$), and overall fraction of edges that join two different communities ($l_{bd}/L$).}
    \label{fig:boundary_boxplot}
\end{figure*}

The average values of $F_{bd}$, $l_{bd}/L$, and $\ln BF$ are tabulated in Table \ref{tab:boundarytable}, while their distributions are illustrated in Figure \ref{fig:boundary_boxplot}. The Bayes factors are lowest for C99999P00001 and C92P08. Moreover, the spread of $F_{bd}$ values for C99999P00001 and C92P08 are the largest, with some structures having very few failures at boundaries in the case of C99999P00001. We observe that models with high compliance weights and no stability objective contain a greater number of vertical beams and are less disordered in structure, which can result in community detection being less useful at characterizing failure locations. Overall, we find that all Bayes factors support the significance of the hypothesis that probability of failure occurring at a boundary is not the same as the probability that a link represents a boundary. 

\begin{table}[]
    \centering
    \begin{tabular}{r|c|c|c}
         Set & $F_{bd}$ & $l_{bd}/L$ & $\ln BF$ \\ \hline
         C99999P00001 & 0.359 & 0.255 & 12.6 \\
         C99P01 & 0.469 & 0.264 & 25.9 \\
         C92P08 & 0.517 & 0.265 & 19.6 \\
         C50S50 & 0.724 & 0.321 & 43.0 \\
         C65S35 & 0.733 & 0.324 & 43.6 \\
         C85P05S10 & 0.576 & 0.279 & 49.6 \\
         C88P01S11 & 0.722 & 0.283 & 96.0 \\
    \end{tabular}
    \caption{Fraction of failures that occur at boundaries between communities ($F_{bd}$), and overall fraction of edges that join two different communities ($l_{bd}/L$). Logarithm of Bayes factor $>$ 5 indicates statistical significance.}
    \label{tab:boundarytable}
\end{table}

\section{Discussion}
We use multi-objective topology optimization to generate networked structures inspired by trabecular bone. An analysis of the stress distribution and fracture patterns in these structures reveals the contribution of compliance, perimeter, and stability objectives to strength and resilience. We observe that in structures with the greatest weight maximizing stiffness, with little to no consideration given to optimizing for stability, mechanical response is sensitive to small geometric perturbations. In comparison, structures generated with greater weight given to the stability objective are more robust. 
       
Each topology-optimized structure analyzed in this paper is constrained to have the same area fraction, but mechanical response can vary widely among structures that otherwise have the same objective weights. This corroborates previous findings that bone mass density is an incomplete predictor of fracture resistance in trabecular bone \cite{mcdonnell_et_al,goldstein1993,goulet1994,brandi2009,fields2009}. Moreover, this variation is most notable for structures optimized primarily for compliance. Prior studies of topology-optimized structures inspired by trabecular bone involve solely compliance minimization with perimeter constraints \cite{boyle_kim_topopt,jang_kim_topopt}. Here, we find that when perimeter and stability weights are taken into account, the reaction force and displacement maxima shift significantly. This may suggest that compliance minimization alone overestimates the behavior of a realistic biological material. Since these materials are typically multifunctional, introducing multiple objectives beyond compliance in topology optimization will provide more flexiblity in balancing various tradeoffs without greatly compromising the mechanical response. When considered on its own as a design principle, Wolff's law, which states that bone adapts itself to resist the loads under which it is placed, and hence typically results in increased bone mass along principal loading axes, may result in structures that are less robust. In real biological tissues, Wolff's law is likely not the sole factor governing remodeling processes, and it may hence be important to use robustness as an objective for bio-inspired design. 

Our mechanical simulations are linearly elastic, followed by brittle failure initiated by a stress-based criterion. An entire beam fails at once when the stress in the beam reaches a specified threshold, but in bone, the nonuniform thicknesses of trabeculae would result in beams that fail progressively. Our division of each beam into five segments serves to mitigate this discrepancy. Moreover, taking into account inelasticity and subscale energy dissipation mechanisms can improve realistic modeling of bone-like structures.

Our observation of substantial variation in the distribution of stress across different models suggests an investigation into the extent to which topology optimization can engineer redundancy in structures. A structure with redundant or sacrificial beams may have higher toughness as the failure of some beams might not immediately result in catastrophic system failure, and stress can be redistributed through remaining beams. 

It will be valuable to draw further biological inspiration from the changes in bone structure that occur due to aging. As bone ages, trabecular architecture increases in anisotropy; trabeculae that are transverse to the principal loading direction are preferentially resorbed, and those that are parallel become thicker \cite{mcdonnell_et_al,thomsen2002}. Currently, our topology-optimization results are static and the objectives used are not chosen with regard to a material that undergoes age-related geometric changes. Additional insight into aging processes can be achieved by extending the modeling procedure to begin with our original topology-optimized structures as initial conditions, followed by an optimization process that reflects the conditions of aging bone.

Overall, the modeling framework developed in this paper has wide-ranging applications for the design of materials and networked structures inspired by nature. While we focus on macroscale architecture in this work, engineering additional architecture at micro- and nanoscales can lead to improved function as bone, along with other naturally-occurring materials, exhibits structure and mechanisms of strength at a range of scales \cite{rho_hierarchical_materials,fratzl_hierarchical_materials}. At the microscale, bone tissue is composed of mineralized collagen fibrils embedded in an organic matrix, and the fibrils themselves comprise mineralized platelets staggered in a regular pattern within a collagen matrix \cite{jager2000}. Other naturally-occurring materials such as nacre contain a similar architecture of elongated platelets organized periodically in a matrix \cite{espinosa_nacre}. Characterizing the contribution of multiscale organization to emergent strength can further inform the development of bio-inspired materials.

\section*{acknowledgements}
We thank Avik Mondal for contributions to the network modeling framework. This work was supported by the David and Lucile Packard Foundation, the Institute of Collaborative Biotechnologies through Army Research Office Grant W911NF-09-D-0001, and the National Science Foundation under Grants EAR-1345074 and CMMI-1435920. Use was made of computational facilities purchased with funds from the National Science Foundation (CNS-1725797) and administered by the Center for Scientific Computing (CSC). The CSC is supported by the California NanoSystems Institute and the Materials Research Science and Engineering Center (MRSEC; NSF DMR 1720256) at UC Santa Barbara. The content of the information does not necessarily reflect the position or the policy of the U.S. government, and no official endorsement should be inferred.

%

\pagebreak
\widetext
\begin{center}
\textbf{\large Supplemental Material}
\end{center}
\setcounter{equation}{0}
\setcounter{figure}{0}
\setcounter{table}{0}
\setcounter{page}{1}
\makeatletter
\renewcommand{\theequation}{S\arabic{equation}}
\renewcommand{\thefigure}{S\arabic{figure}}
\renewcommand{\bibnumfmt}[1]{[S#1]}
\renewcommand{\citenumfont}[1]{S#1}

\section{Methodological considerations}
The individual objective functions in the optimization problem (Eq. 6 of the main text) are normalized so that their respective magnitudes are more consistent. Given that the compliance and stability functions can take values at or near infinity (for purely void structures) and the maximum value of the perimeter function is limited by the filter length scale (an alternating solid-void design with no structural links), a finite normalization scheme may disproportionately normalize those functions relative to the perimeter function. As a result the function weights may appear disproportionate, but could be made more similar (or even uniform) by using modified normalization factors. For reference, Table \ref{tab:Function_contributions} shows the percent of the weighted objective sum contributed by each function for the compliance and perimeter models (a 50-50 split indicates equal weighted values for each function).

\begin{table}[]
    \centering
    \begin{tabular}{|l|c|c|}
        \hline
         & Compliance & Perimeter \\
        Model & Contribution & Contribution \\
        \hline
        C92P08\_1 & 63.2 & 36.8 \\
        C92P08\_2 & 64.3 & 35.7 \\
        C92P08\_3 & 65.0 & 35.0 \\
        C92P08\_4 & 63.2 & 36.8 \\
        C92P08\_5 & 77.9 & 22.1 \\
        C92P08\_6 & 68.4 & 31.6 \\
        C92P08\_7 & 61.3 & 38.7 \\
        C92P08\_8 & 72.0 & 28.0 \\
        C92P08\_9 & 70.5 & 29.5 \\
        C92P08\_10 & 56.5 & 43.5 \\
        C92P08\_11 & 62.4 & 37.6 \\
        C92P08\_12 & 65.8 & 34.2 \\
        C99999P00001\_1 & 100.0 & 0.0222 \\
        C99999P00001\_2 & 100.0 & 0.0288 \\
        C99999P00001\_3 & 100.0 & 0.0303 \\
        C99999P00001\_4 & 100.0 & 0.0274 \\
        C99999P00001\_5 & 100.0 & 0.0192 \\
        C99999P00001\_6 & 100.0 & 0.0118 \\
        C99999P00001\_7 & 100.0 & 0.0083 \\
        C99999P00001\_8 & 100.0 & 0.0158 \\
        C99999P00001\_9 & 100.0 & 0.0117 \\
        C99999P00001\_10 & 100.0 & 0.0140 \\
        C99999P00001\_11 & 100.0 & 0.0220 \\
        C99999P00001\_12 & 100.0 & 0.0205 \\
        C99P01\_1 & 94.6 & 5.4 \\
        C99P01\_2 & 90.7 & 9.3 \\
        C99P01\_3 & 92.2 & 7.8 \\
        C99P01\_4 & 91.5 & 8.5 \\
        C99P01\_5 & 93.9 & 6.1 \\
        C99P01\_6 & 94.3 & 5.7 \\
        C99P01\_7 & 96.0 & 4.0 \\
        C99P01\_8 & 93.9 & 6.1 \\
        C99P01\_9 & 95.0 & 5.0 \\
        C99P01\_10 & 93.0 & 7.0 \\
        C99P01\_11 & 95.0 & 5.0 \\
        C99P01\_12 & 93.7 & 6.3 \\
        \hline
    \end{tabular}
    \caption{Percent contribution of compliance and perimeter functions to weighted objective sum for models with no stability objective.}
    \label{tab:Function_contributions}
\end{table}

\section{Topology-optimized structures}
\begin{figure*}
    \centering
    \includegraphics[width=\linewidth]{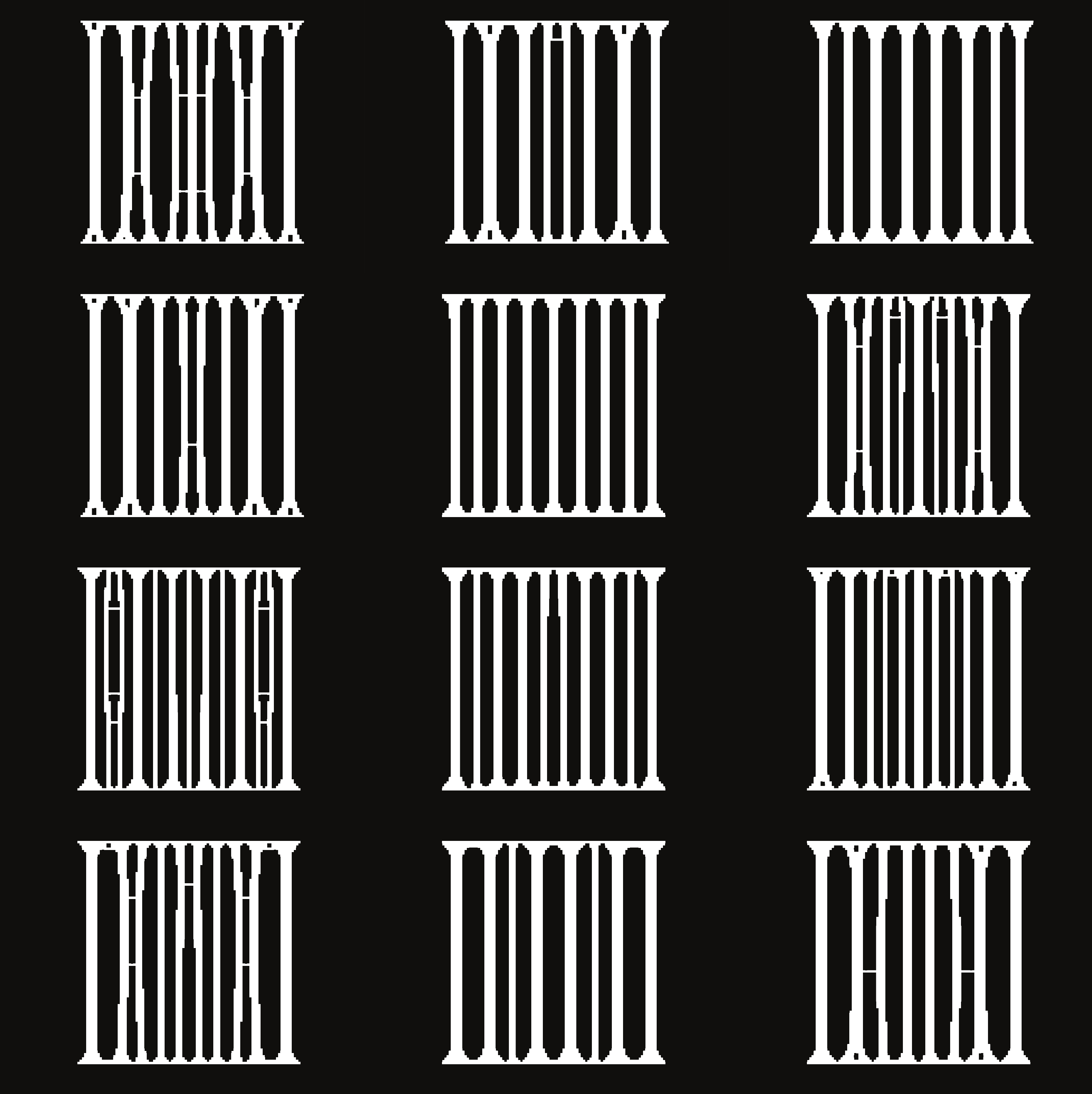}
    \caption{Structures generated using C99999P00001 objective weights.}
    \label{fig:C99999}
\end{figure*}

\begin{figure*}
    \centering
    \includegraphics[width=\linewidth]{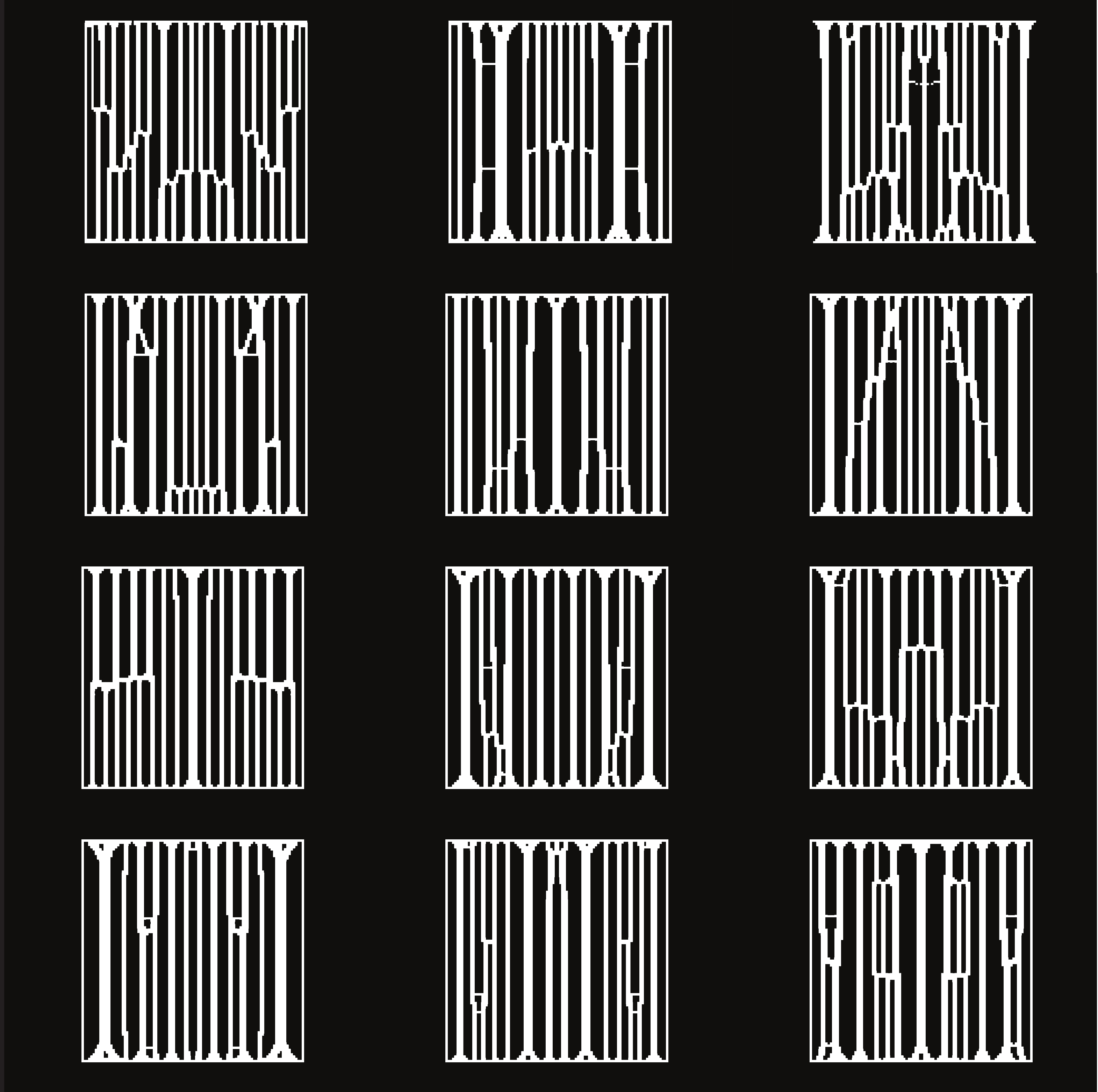}
    \caption{Structures generated using C99P01 objective weights.}
    \label{fig:C99}
\end{figure*}

\begin{figure*}
    \centering
    \includegraphics[width=\linewidth]{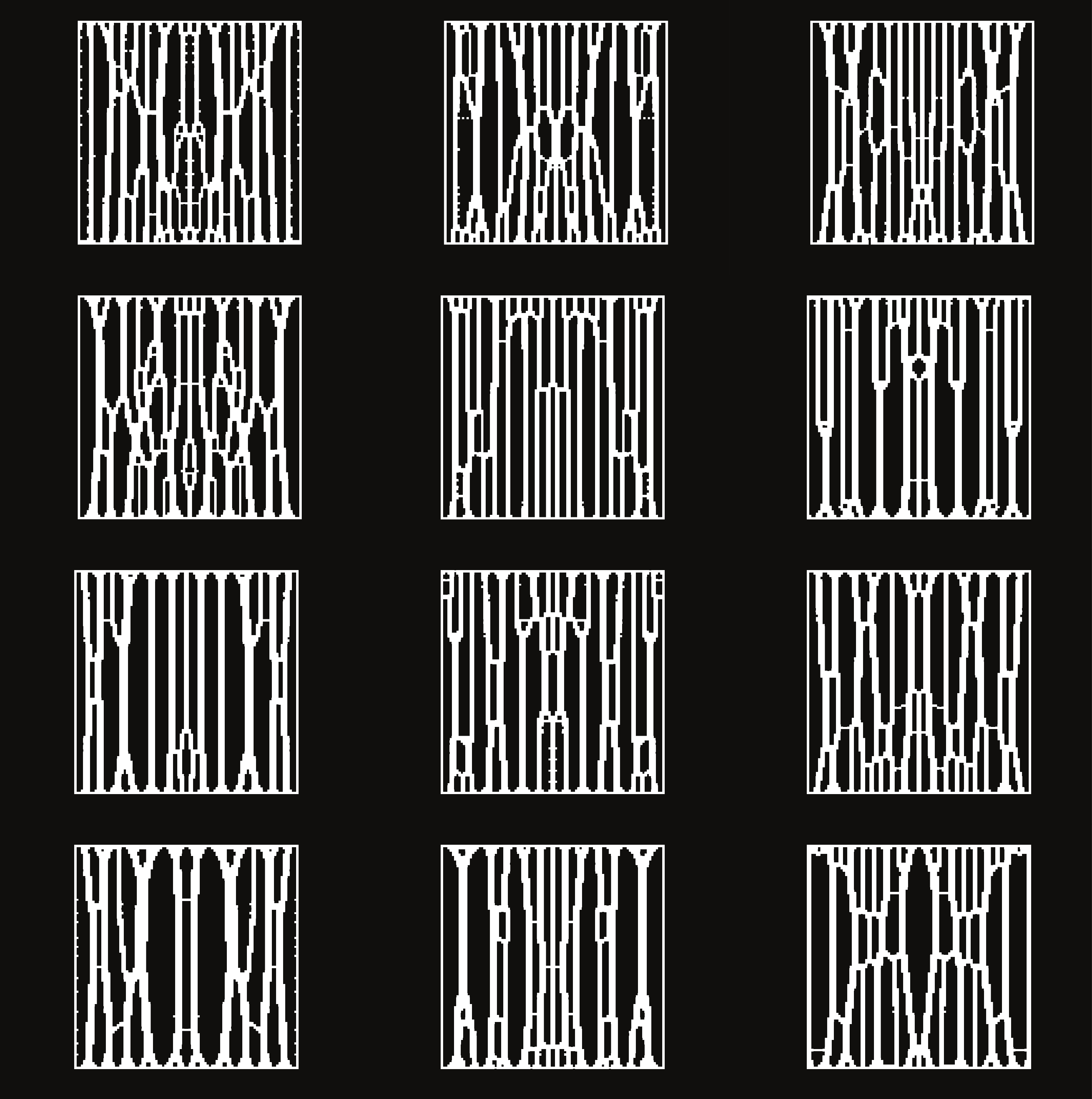}
    \caption{Structures generated using C92P08 objective weights.}
    \label{fig:C92}
\end{figure*}

\begin{figure*}
    \centering
    \includegraphics[width=\linewidth]{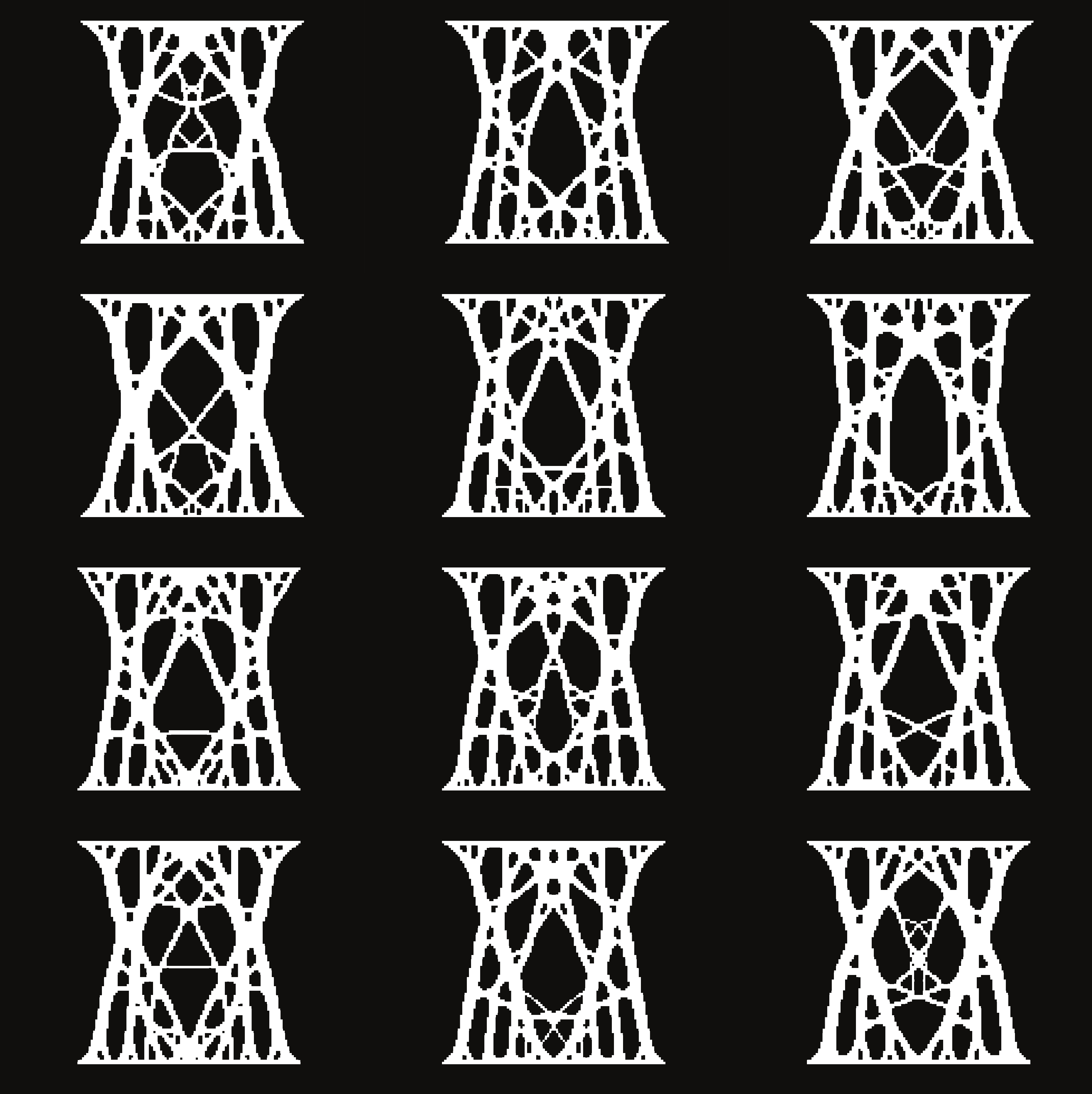}
    \caption{Structures generated using C50S50 objective weights.}
    \label{fig:C50}
\end{figure*}

\begin{figure*}
    \centering
    \includegraphics[width=\linewidth]{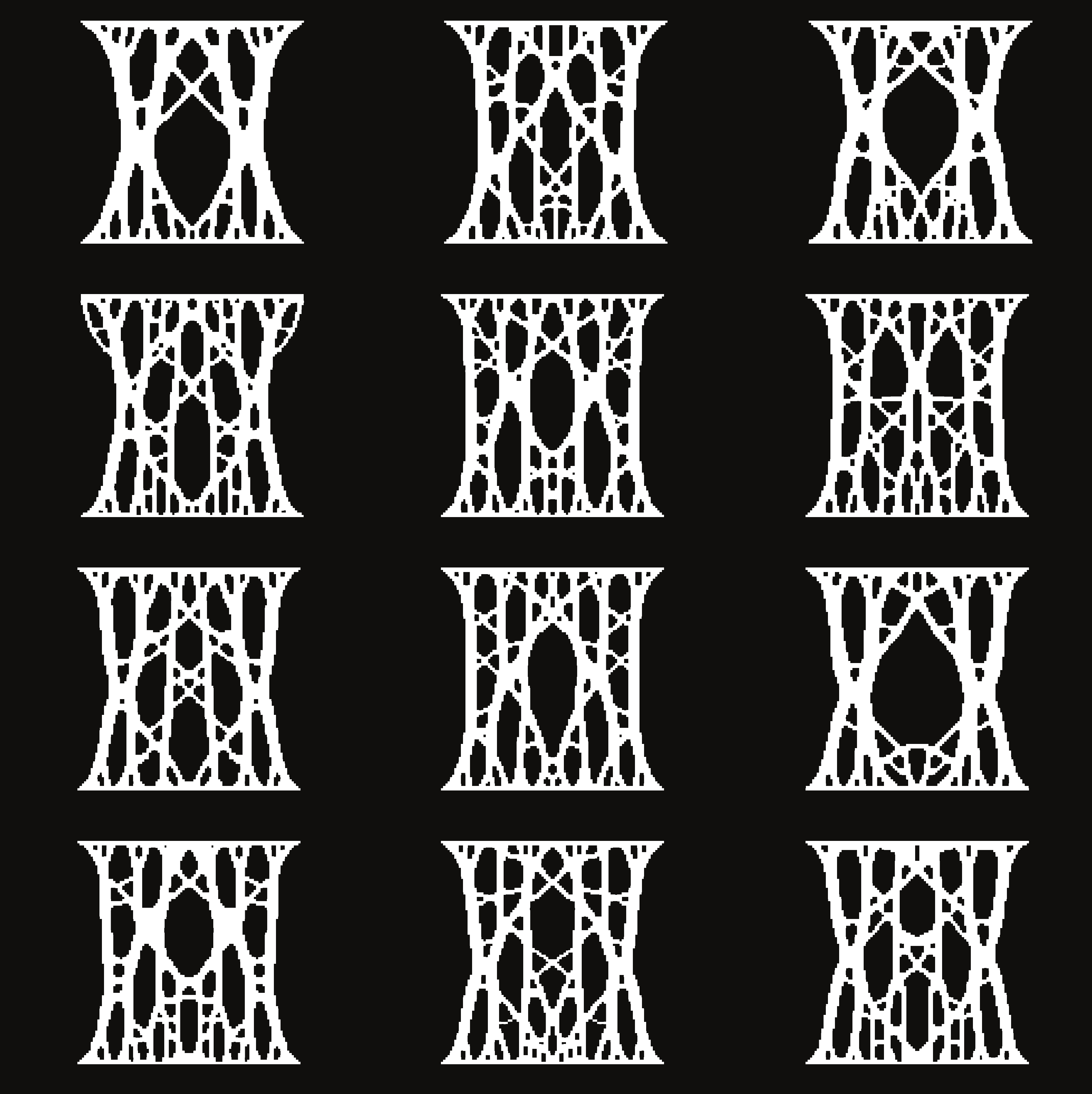}
    \caption{Structures generated using C65S35 objective weights.}
    \label{fig:C65}
\end{figure*}

\begin{figure*}
    \centering
    \includegraphics[width=\linewidth]{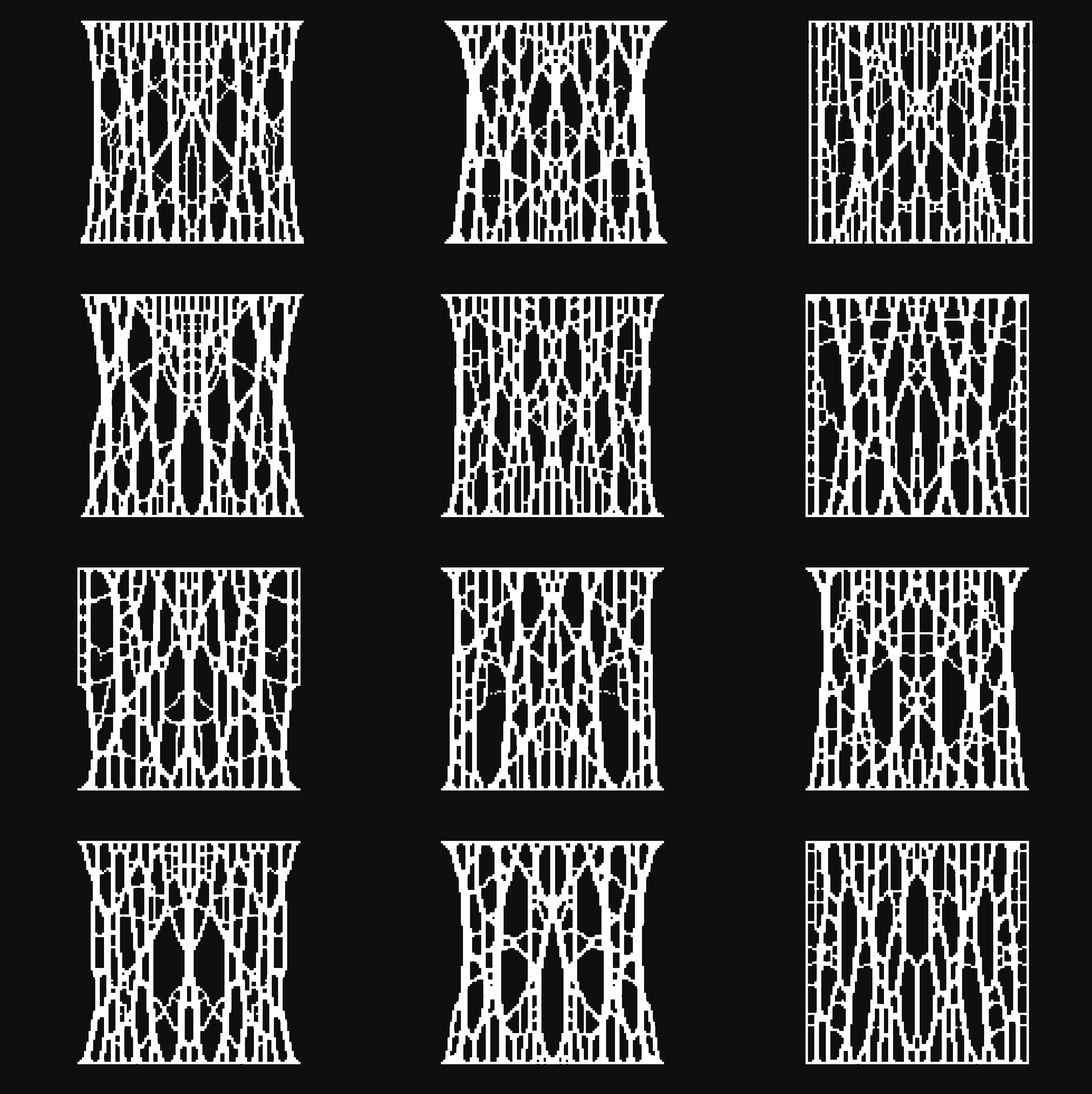}
    \caption{Structures generated using C85P05S10 objective weights.}
    \label{fig:C85}
\end{figure*}

\begin{figure*}
    \centering
    \includegraphics[width=\linewidth]{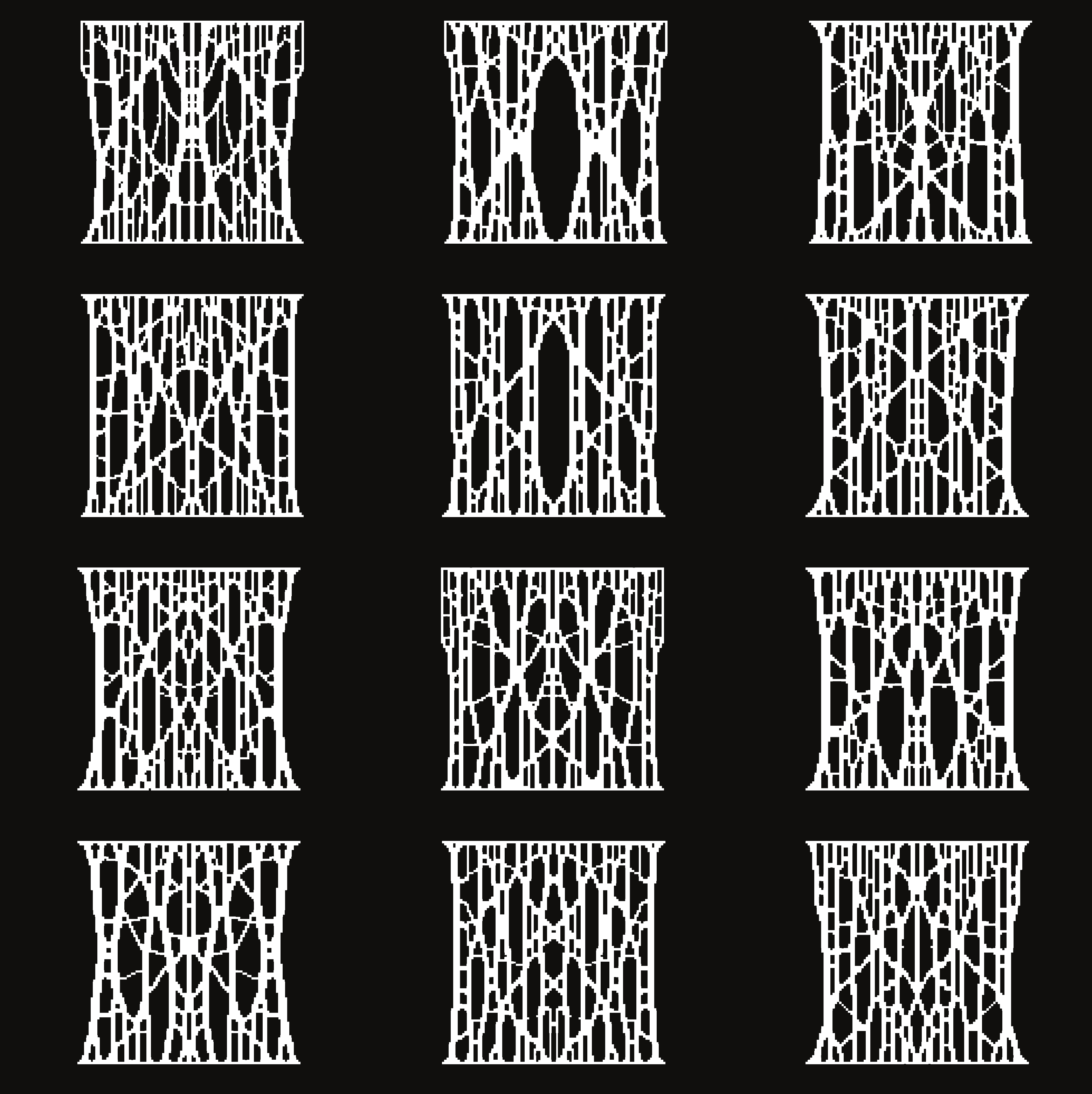}
    \caption{Structures generated using C88P01S11 objective weights.}
    \label{fig:C88}
\end{figure*}

Twelve structures were generated for each of the seven sets of objective weights. The initial density distribution was perturbed slightly for each optimization run to produce variation in architecture. Structures for each parameter set are shown in Figs. \ref{fig:C99999}-\ref{fig:C88}.


\section{Force-displacement response}
While we only consider the force-displacement response between the origin and the point at which it has reached zero for each structure (indicating total failure), we include the full force-displacement curves here for completeness (Fig. \ref{fig:forcedisp}). The data used to generate Fig. 4 of the main text are also shown as boxplots in Fig. \ref{fig:boxplot} to facilitate comparison between the response of the original and perturbed models.

\begin{figure*}
    \centering
    \includegraphics[width=\linewidth]{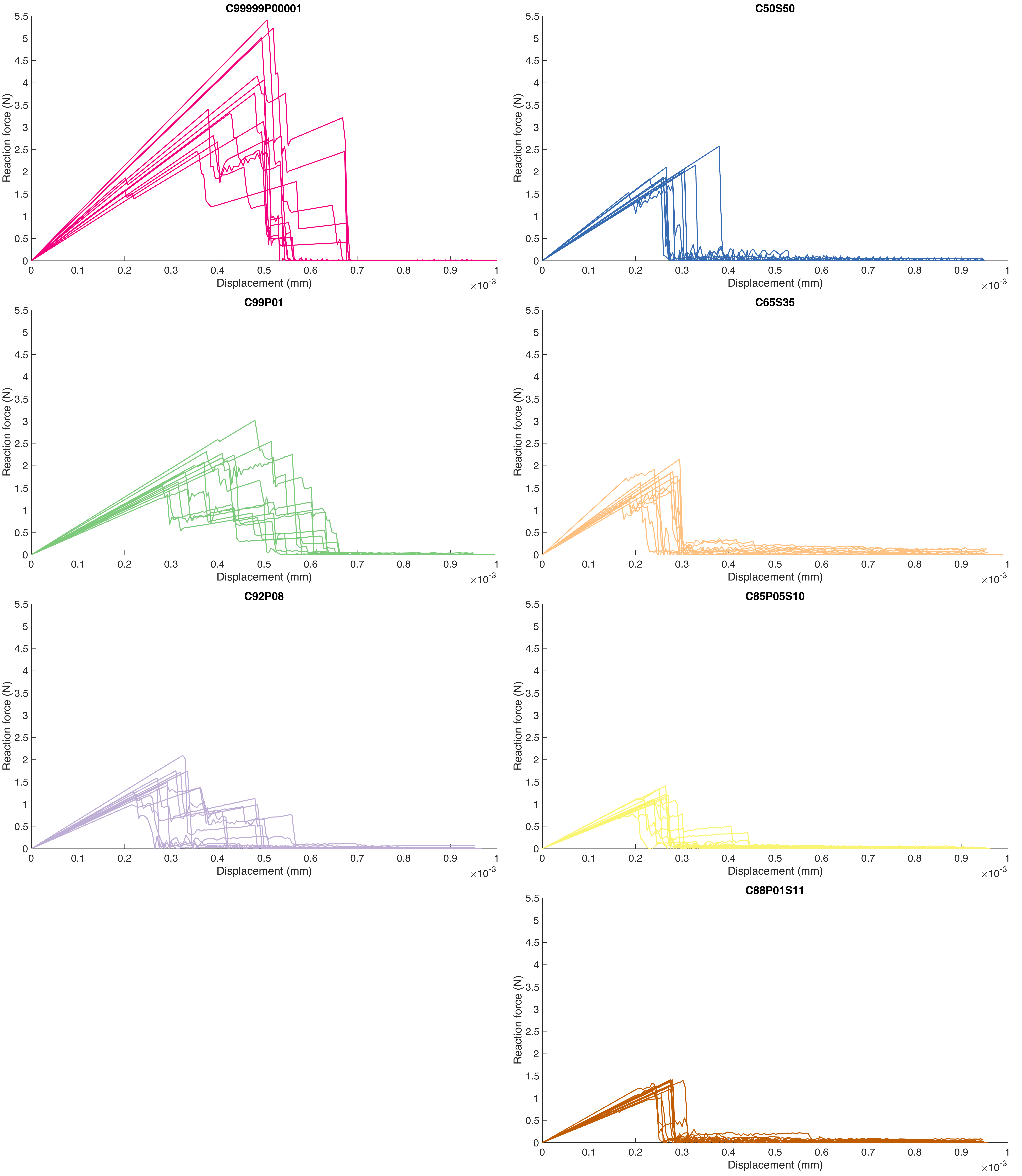}
    \caption{Force-displacement curves for each structure in a parameter set.}
    \label{fig:forcedisp}
\end{figure*}

\begin{figure*}[]
\centering
    \includegraphics[width=\textwidth]{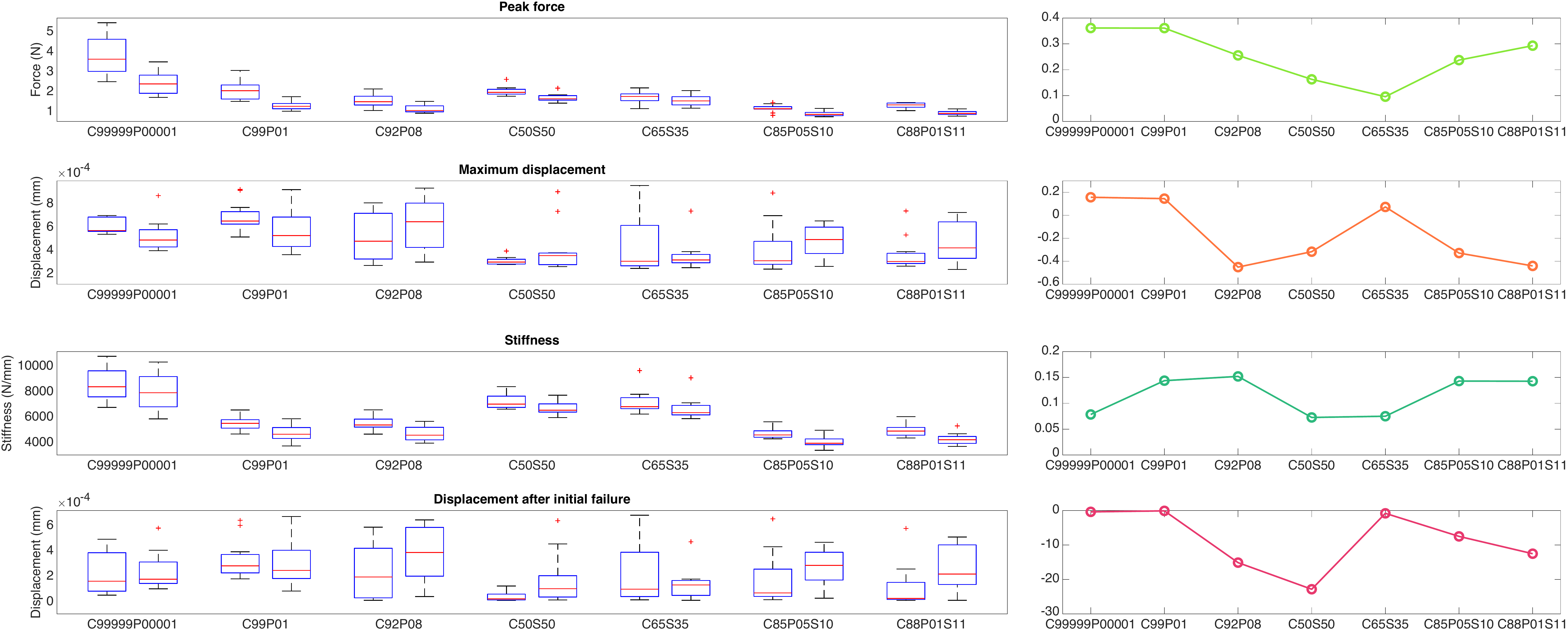}
    \caption{Left column: boxplots indicating variation in peak force, maximum displacement, stiffness, and displacement between initial beam failure and system failure. Each pair of plots represents the same parameter set, with the left boxplot corresponding to the original model and the right boxplot to the perturbed model.  Right column: percent difference between original and perturbed model, averaged over each model within the same parameter set, for each of the four metrics shown on the left. }
    \label{fig:boxplot}
    \end{figure*}

\end{document}